\documentclass[twocolumn,showpacs,showkeys,amsmath,amssymb,prl]{revtex4} 

\usepackage{graphicx}   
\usepackage{dcolumn}    
\usepackage{bm}         
\begin{document} 
 
\title{Differences between mean-field dynamics and $N$-particle quantum dynamics  as a signature of entanglement} 
 
\author{Christoph Weiss$^1$}
\email{weiss@theorie.physik.uni-oldenburg.de}
\author{Niklas Teichmann$^2$}

\affiliation{$^1$Laboratoire Kastler Brossel, \'Ecole Normale Sup\'erieure, Universit\'e Pierre et Marie-Curie-Paris 6, 24 rue Lhomond, CNRS,
                F-75231 Paris Cedex 05, France
} 

   \affiliation{$^2$Institut Henri Poincar\'e, Centre Emile Borel, 11 rue P.\ et M.\ Curie, F-75231 Paris Cedex 05, France}

                  
\date{Submitted: 27 July 2007, published 10 April 2008}
 
\begin{abstract}
A Bose-Einstein condensate in a tilted double-well potential under the influence of time-periodic potential differences is investigated in the regime where the mean-field (Gross-Pitaevskii) dynamics become chaotic. For some parameters near stable regions, even averaging over several condensate oscillations does not remove the differences between mean-field and $N$-particle results. While introducing decoherence via piecewise deterministic processes reduces those differences, they are due to the emergence of mesoscopic entangled states in the chaotic regime.
\end{abstract} 


\maketitle

Experimentally it is possible to generate precisely controllable double-well potentials for Bose-Einstein condensates (BECs) (Ref.~\cite{GatiOberthaler07} and references therein). A future goal for this system is the realization of mesoscopic entanglement~\cite{GatiOberthaler07}. When combined with a time-periodic potential difference between the two wells, a BEC in a double well could also be used to investigate quantum chaos~\nocite{UtermannEtAl94}\nocite{AbdullaevKraenkel00}\nocite{GhoseEtAl01}\nocite{LeeEtAl01}\cite{UtermannEtAl94,GhoseEtAl01,LeeEtAl01,AbdullaevKraenkel00}. 
Another system which is widely used to investigate quantum chaos is the quantum delta-kicked rotor~\cite{Moore95,dArcy01,CreffieldEtAl06}.
 Research on quantum chaos includes topics like quantum signatures of chaos~\cite{Haake92}, quasi-stationary distributions~\cite{BreuerEtAl00}, entanglement~\cite{Garcia-mata07,GhoseSanders04} and decoherence~\cite{Braun01}.

Often,  a mean-field approach within the Gross-Pitaevskii equation is applied to describe BECs. Still, there are noticeable differences between
 mean-field  dynamics and quantum dynamics: only the latter displays the well known collapse and revival phenomenon (cf.~\cite{GreinerEtAl02b}). By time-averaging over several of those oscillations, these differences usually disappear.
However, preliminary results~\cite{TeichmannEtAl06}  for the periodically driven double-well potential indicate that even under time-average, mean-field dynamics and quantum dynamics can display qualitatively different results in  the regime for which the mean-field dynamics become chaotic.

In this Letter, these differences are investigated systematically. First, the  $N$-particle
Hamiltonian is introduced for which the Gross-Pitaevskii equation corresponds to a driven
nonrigid pendulum. If decoherence is implemented on the $N$-particle level via piecewise
deterministic processes, the quantum dynamics can become qualitatively similar to the
mean-field dynamics. The reason for the remaining differences between both approaches is the emergence of mesoscopic entangled states.

To describe a BEC in a double well with single-particle tunneling frequency~$\Omega$ and pair
interaction energy~$2\hbar\kappa$,  we use the Hamiltonian in two-mode
approximation~\cite{MilburnEtAl97}:
\begin{eqnarray}
\label{eq:H}
\hat{H} &=& -\frac{\hbar\Omega}2\left(\hat{a}_1^{\phantom\dag}\hat{a}_2^{\dag}+\hat{a}_1^{\dag}\hat{a}_2^{\phantom\dag} \right) + \hbar\kappa\left(\hat{a}_1^{\dag}\hat{a}_1^{\dag}\hat{a}_1^{\phantom\dag}\hat{a}_1^{\phantom\dag}+\hat{a}_2^{\dag}\hat{a}_2^{\dag}\hat{a}_2^{\phantom\dag}\hat{a}_2^{\phantom\dag}\right)\nonumber\\
&+&\hbar\big(\mu_0+\mu_1\sin(\omega t)\big)\left(\hat{a}_2^{\dag}\hat{a}_2^{\phantom\dag}-\hat{a}_1^{\dag}\hat{a}_1^{\phantom\dag}\right)\;,
\end{eqnarray}
where $\hat{a}^{(\dag)}_j$ creates (annihilates) a boson in well~$j$; $\mu_0$ models the tilt and $\mu_1$ is the driving amplitude. Such Hamiltonians have been used for schemes of entanglement generation~\cite{TeichmannWeiss07,Creffield2007}; 
without the periodic driving, entanglement has been investigated in BECs~\cite{MicheliEtAl03,MahmudEtAl03}. Other applications include high precision measurements, many-body quantum coherence~\cite{Lee06,Lee07} and spin systems~\cite{DusuelVidal05}.

\begin{figure}
\includegraphics[angle=-90,width=\linewidth]{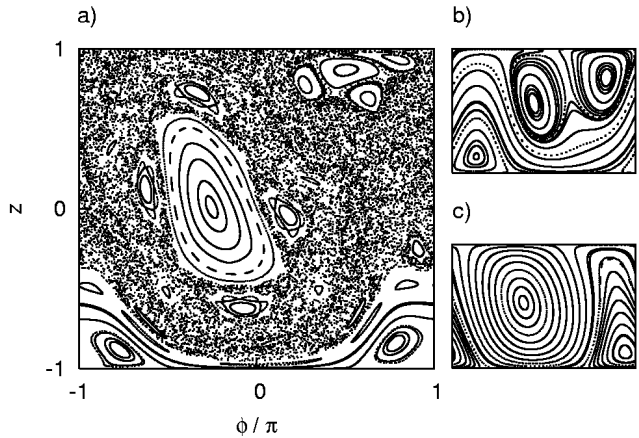}
\caption{Poincar\'e surface of section for the forced nonrigid pendulum (the mean-field dynamics~(\ref{eq:mean}) are plotted for various starting points at integer multiples of the oscillation period~$2\pi/\omega$). Closed loops are characteristic for stable orbits whereas irregular dots represent chaotic regions. For a BEC in a double-well, the parameters correspond to: \\
{(a)} a tilt of $2\mu_0/\Omega=3.0$, a driving frequency of~$\omega=3\Omega$, an interaction of $N\kappa/\Omega=0.8$ and a driving amplitude of $2\mu_1/\Omega=0.9$ (i.e. a one-``photon''-resonance~\cite{EckardtEtAl05}),
\\
{(b)} the $3/2$-``photon''-resonance  with $N\kappa/\Omega=0.1$,  $2\mu_0/\Omega=3.0$, $\omega/\Omega=2.08$ and~$2\mu_1/\Omega=1.8$,\\
{(c)} all parameters as in a) except for $N\kappa/\Omega=0.3$.
}
\label{fig:poin}
\end{figure}

 On the level of the Gross-Pitaevskii equation for the above model, a wave function is characterized by 
the variables $\theta$ and $\phi$, where $\cos^2[\theta/2]$ ($\sin^2[\theta/2]$) is the probability of finding the condensate in well~1 (well~2) and $\exp(i\phi)$ is the phase between the two wells. The corresponding $N$-particle wave-function (``atomic coherent states''~\cite{MandelWolf95}) with all particles in this state reads (in an expansion in the Fock-basis $| n, N-n \rangle$ with $n$ atoms in well~$1$):
\begin{eqnarray}
\label{eq:atomic}
\left|\theta,\phi\right>&=& \sum_{n=0}^N \genfrac{(}{)}{0pt}{}{N}{n}^{1/2}\cos^{n}(\theta/2)
                 \sin^{N-n}(\theta/2)
                 \nonumber\\
                 &\times& e^{i(N-n)\phi}| n, N-n \rangle\;.
\end{eqnarray}
The mean-field dynamics can be mapped to that of a nonrigid pendulum~\cite{SmerziEtAl97,TeichmannEtAl06}; including periodic driving the 
Hamilton function reads ($z=\cos^2(\theta/2)-\sin^2(\theta/2)$):
\begin{eqnarray}
H_{\rm mf}& = &\frac{N\kappa}{\Omega}z^2-\sqrt{1-z^2}\cos(\phi)\nonumber\\
&-&2z\left(\frac{\mu_0}{\Omega}+
\frac{\mu_1}{\Omega}\sin\left({\textstyle\frac{\omega}{\Omega}}\tau\right)\right)\;,\quad \tau =t\Omega\;.
\label{eq:mean}
\end{eqnarray}
The experimentally measurable~\cite{GatiOberthaler07} population imbalance  $z/2$ can be used
to characterize the mean-field dynamics. 
Fig.~\ref{fig:poin} shows typical Poincar\'e surfaces of section. The initial parameters were
chosen such that tunneling in the driven, tilted double-well potential is enhanced by
``photon''-assisted tunneling~\cite{EckardtEtAl05} (cf.\ Ref.~\cite{GrifoniHanggi98}).  If
the interaction is not too low ($N\kappa/\Omega \gtrapprox 0.4\ldots0.6$), regular and
chaotic dynamics coexist (Fig.~\ref{fig:poin}.a \textit{cf.}~\cite{GuckenheimerHolmes83}), for low interaction the dynamics are regular (Figs.~\ref{fig:poin}.b and~\ref{fig:poin}.c).

\begin{figure}
\includegraphics[angle=-0,width=\linewidth]{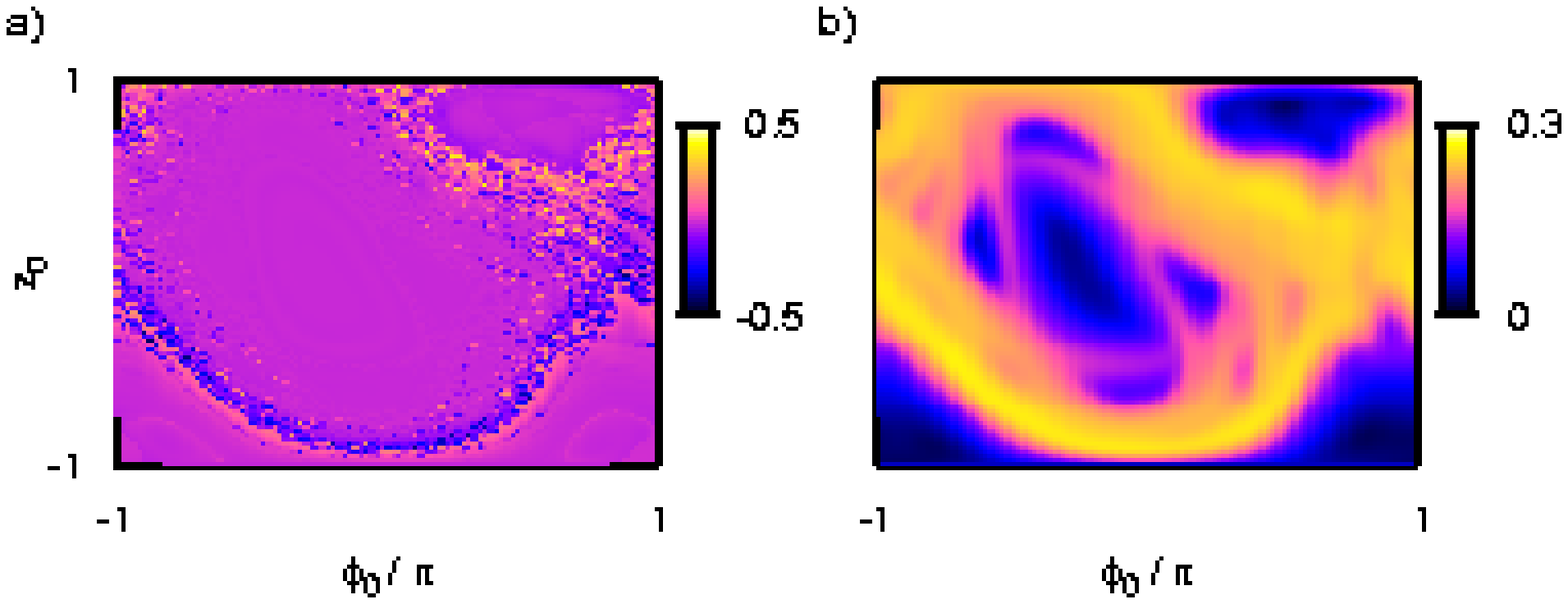}
\caption{(color online) Quantum dynamics ($N=100$) versus mean-field dynamics using the parameters of Fig.~\ref{fig:poin}.a. \\
(a) The difference of the time-averaged population imbalances  $\langle J_z\rangle_T/N$ and $\langle z/2\rangle_T$ as a function of $101^2$ initial conditions ($z_0$, $\phi_0$) in a two-dimensional projection of the resulting three-dimensional plot ($T=100/\Omega$).
\\ (b) The time-averaged root-mean-square (r.m.s.)-fluctuations $\langle\Delta J_z\rangle_T/N$ of the population imbalance as a function of the initial atomic coherent state~(\ref{eq:atomic}). }
\label{fig:nlssgl}
\end{figure}

For the parameters corresponding to the Poincar\'e surface of section in Fig.~\ref{fig:poin}.a, Fig.~\ref{fig:nlssgl}.a displays the differences between $N$-particle and mean-field dynamics by numerically calculating (using the Shampine-Gordon-routine~\cite{ShampineGordon75}) the  time-average of the (experimentally measurable~\cite{GatiOberthaler07}) population imbalance $\langle J_z\rangle/N$ (which corresponds to the mean-field $z/2$):
\begin{equation}
\frac{\langle J_z\rangle_T}N=\frac1{NT}\int_0^Tdt\frac12\langle\psi|\hat{a}_1^{\dag}\hat{a}_1^{\phantom\dag}-\hat{a}_2^{\dag}\hat{a}_2^{\phantom\dag}|\psi\rangle\;,
\end{equation} 
where for $\langle J_z\rangle/N=\pm 0.5$ the entire condensate is in the left, respectively, right well. Each point represents an initial condition~(\ref{eq:atomic}). The differences are small if the mean-field dynamics are regular (cf. Fig.~\ref{fig:poin}.a) while they can be rather large in the chaotic regime (up to half the theoretical limit, ${\rm max}\{|z/2-\langle J_z\rangle/N|\}=1$). Most of the deviations between $N$-particle dynamics and mean-field dynamics in Fig~\ref{fig:nlssgl}.a lie within twice the root-mean-square (r.m.s.)-fluctuations of the $N$-particle dynamics.  However, contrary to the preliminary results of Ref.~\cite{TeichmannEtAl06}, for many initial conditions in the (classically) chaotic regime the differences can be very small; they are large near the boundaries of stable regions.

In Fig.~\ref{fig:nlssgl}.b, the time-averaged r.m.s.-fluctuations of $\langle J_z\rangle/N$ reproduce many features displayed in the Poincar\'e section in Fig.~\ref{fig:poin}.a. 
Note that the values for the r.m.s.-fluctuations are well above those expected for $N=100$ particles in an atomic coherent state, $\sin(\theta)/(2\sqrt{N})\leq 0.05$, thus clearly indicating that more than one atomic coherent state is involved. Bose-Einstein condensates of $N\approx 100$ have been realized experimentally~\cite{ChuuEtAl05}, both the validity of the two-mode approximation will be better and life-times of mesoscopic entangled states will be longer than in larger condensates. However, even when the calculation is repeated for  $N=1000$ particles, the differences in the chaotic regime remain. As the (non-linear) Gross-Pitaevskii equation does not allow any superpositions, decoherence should reduce the differences between mean-field and quantum dynamics.

\begin{figure} 
\includegraphics[angle=-90,width=\linewidth]{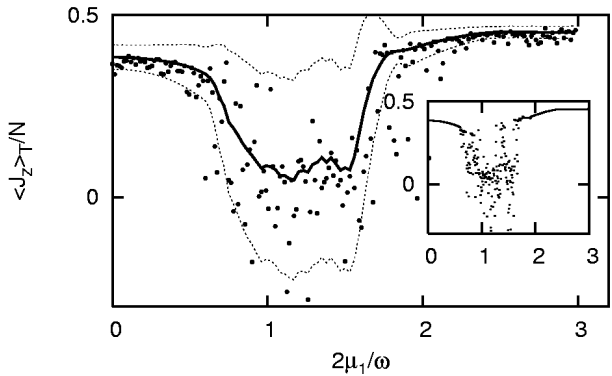}
\caption{Time-averaged population imbalance  $\langle J_z\rangle_T/N$  for various driving amplitudes~$\mu_1$ in a tilted driven double well ($2\mu_0/\Omega=3.0$, $\omega=3\Omega$, $T=100/\Omega$). The BEC initially is in the lower well ($z_0=1$). 
Solid line: $\langle J_z\rangle_T/N$ for $N=1000$ is a smooth curve as opposed to the mean-field results depicted in the inset, which display chaotic jumps for small changes of the driving amplitude.
Dots in the main plot: If decoherence is included via the PDP-process described around
Eq.~(\ref{eq:pdphere}) with on average $\simeq 5$ jumps ($\alpha=1/20$)~\cite{footnoteCWNT08}, the behavior is closer to the mean-field dynamics. Many dots lie in the area defined by the curves $(\langle J_z\rangle_T\pm\langle \Delta J_z\rangle_T)/N$ (dashed lines).}
\label{fig:N1000}
\end{figure}

In this Letter, we use a piecewise deterministic process (PDP) (Ref.~\cite{Breuer06}, cf.\ \cite{DalibardEtAl92}) to model decoherence. To avoid to have to introduce decoherence also on the mean-field level (the atomic coherent states~(\ref{eq:atomic}) become orthogonal in the limit~$N\to\infty$), we use the projection on the atomic coherent states~\cite{MandelWolf95}:
\begin{equation}
\label{eq:one}
\mathbf{1}=\frac{N+1}{4\pi}\int d\theta\sin(\theta)\int d\phi \left|\theta,\phi\right>\left<\theta,\phi\right|\;.
\end{equation}
Now, the PDP simplifies to having jumps on one of the atomic coherent states~(\ref{eq:atomic}) after time~$t$ with probability
\begin{equation}
\label{eq:pdphere}
p_{\rm jump} = 1-\exp(-\alpha t)\;,\quad \alpha={\rm const.} >0\;,
\end{equation}
and Hamiltonian dynamics~(\ref{eq:H}) between jumps.
The state on which the wave-function is projected is determined by the
probability distribution~
\begin{equation}
\label{eq:dist}
p_{\rm \theta,\phi}\,d\Omega=\frac{N+1}{4\pi}\left|\langle\psi|\theta,\phi\rangle\right|^2\sin(\theta)\,d\theta\,d\phi\;.
\end{equation}
Figure~\ref{fig:N1000} shows that the PDP can qualitatively reproduce the results of the
Gross-Pitaevskii equation~\cite{footnoteCWNT08}. Without introducing the decoherence, the
qualitative difference between mean-field and quantum dynamics are quite large; averaging
over several PDPs would again result in a smooth curve within the error-bars in
Fig.~\ref{fig:N1000}. While BEC-research in quantum chaos often assumes the validity of the
Gross-Pitaevskii
equation~\cite{UtermannEtAl94,GhoseEtAl01,LeeEtAl01,AbdullaevKraenkel00,MartinEtAl07}, at
least for the model investigated here, only decoherence can lead to the chaotic behavior
predicted by mean field.

\begin{figure}
\includegraphics[angle=0,width=\linewidth]{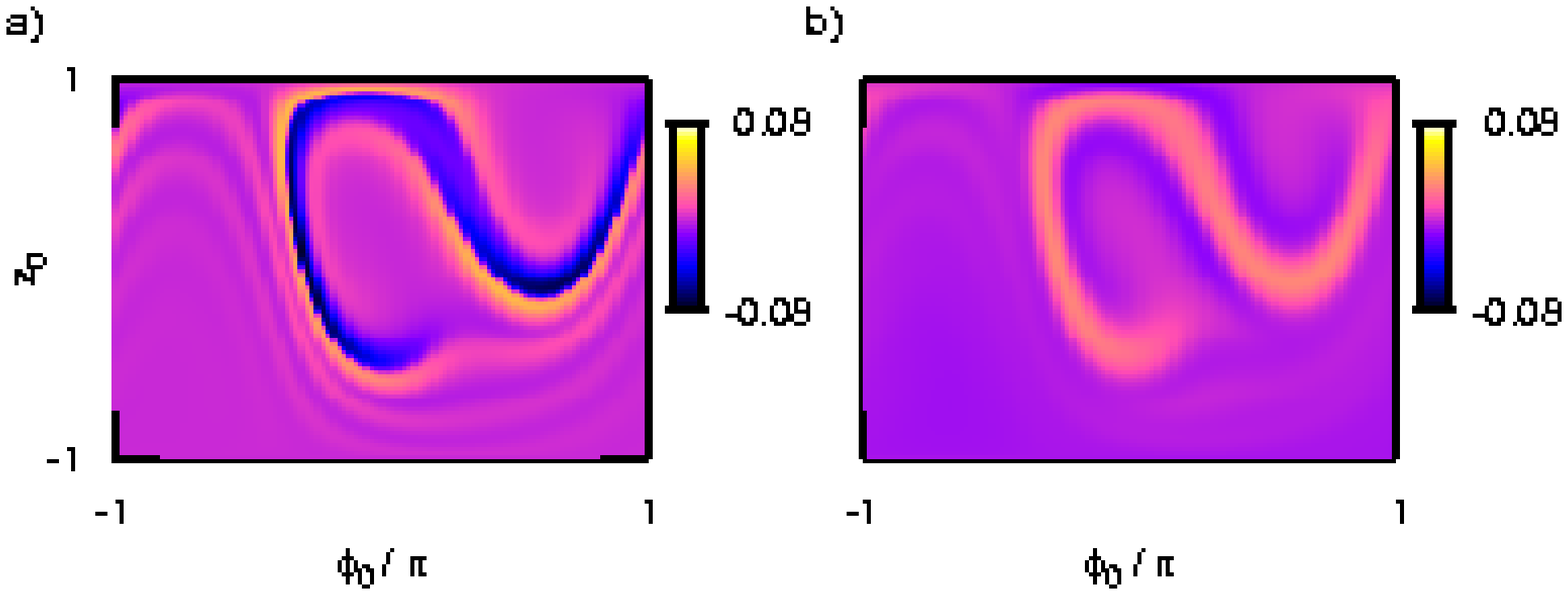}
\caption{(color online)   Time-averaged population imbalances of quantum dynamics ($N=100$) versus mean-field dynamics at the $3/2$-``photon''-resonance  of Fig.~\ref{fig:poin}.b.  
\\
(a) The difference is plotted as a function of the initial condition ($z_0$, $\phi_0$) in a two-dimensional projection ($T=100/\Omega$).  \\
(b) As in a) but the mean-field dynamics are replaced by an average 
over the distribution of initial conditions~(\ref{eq:dist}).
}
\label{fig:dreihalbe}
\end{figure}

Furthermore, 
differences between quantum dynamics  and mean-field dynamics can also occur in the regular regime:
Fig.~\ref{fig:dreihalbe} shows that, at least for $N=100$, the differences can even lie above the result for many initial conditions in the chaotic regime (Fig.~\ref{fig:nlssgl}.a).
One way to reduce the differences is to average over the Husimi distribution~(\ref{eq:dist}) (see Fig.~\ref{fig:dreihalbe}, cf.\ Refs.~\cite{UtermannEtAl94,StrzysEtAl08} and references therein).
This decreases the peaks of the differences between mean-field and quantum dynamics by a factor of 2 (in the chaotic regime, the  factor can be of the order of~5). A perfect agreement cannot be expected as the averaged probability distribution on the mean-field level is always added whereas in quantum mechanics also destructive interference can occur.

\begin{figure}
\includegraphics[width=\linewidth]{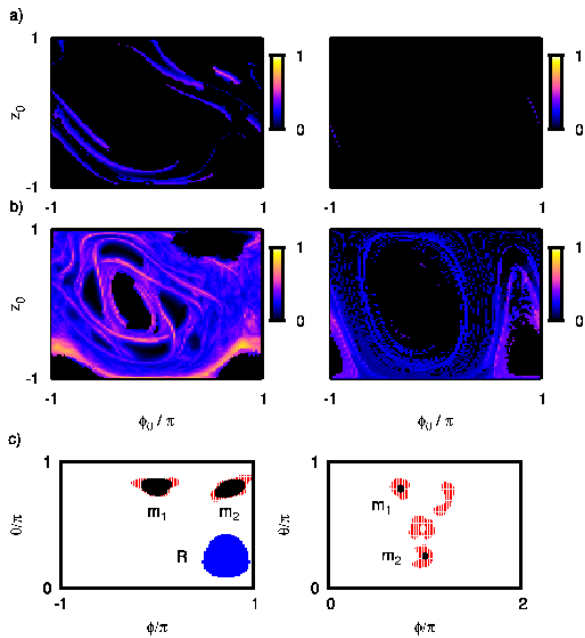}
\caption{(color online) Entanglement~(\ref{eq:ent}) for parameters as in Fig.~\ref{fig:poin}.a (left column) and as in Fig.~\ref{fig:poin}.c (right column).\\ (a), (b): Mesoscopic quantum superpositions were identified at times $\tau=5,10,15,\ldots $; the maximum value of~$\sigma_{\rm ent}$ is displayed for $101^2$ initial conditions ($z_0$, $\phi_0$) and for
 (a) short times ($\tau=10$) and 
(b) longer times ($\tau=100$).\\
(c) Projection of two characteristic entangled states (with maxima $m_1$, $m_2$) on the atomic coherent states~(\ref{eq:atomic}). Black (grey/red) regions: $\left|\langle\theta,\phi|\psi\rangle\right|^2>0.16$ ($>0.05$). Left: $z_0=-0.6$, $\phi_0=-2.764601535$, $\tau=80$, $\sigma_{\rm ent}\simeq 72.3\%$ . Right: $z_0=-0.98$, $\phi_0=-2.701769682$, $\tau=75$, $\sigma_{\rm ent}\simeq 33.5\%$.
In the left plot, the large blue/grey circle is a typical set~R
around~$|\widetilde{\theta},\widetilde{\phi}\rangle$ with
$|\langle\theta,\phi|\widetilde{\theta},\widetilde{\phi}\rangle|^2>0.001$ (cf.\ Eq.~(\ref{eq:fiddef})).}
\label{fig:ent} 
\end{figure}

On the level of quantum dynamics, the differences could be due to either a distribution of many atomic coherent states - or maybe even mesoscopic superpositions.   
For our model all mesoscopic quantum superpositions of all $N$ particles being either in one quantum state or in another can be expressed as a sum of two atomic coherent states (see the explanation before Eq.~(\ref{eq:atomic})):
\begin{equation}
\label{eq:super}
|\psi_{\rm sp}\rangle = \eta\left( |\theta_1,\phi_1\rangle + e^{i\gamma}|\theta_2,\phi_2\rangle\right)\,,\; 0\le\gamma\le 2\pi
\end{equation}
If both parts hardly overlap, $|\langle\theta_1,\phi_1|\theta_2,\phi_2\rangle|\ll 1$, the normalization $\eta\simeq 1/\sqrt{2}$ and $|\psi_{\rm sp}\rangle$ is a highly entangled mesoscopic state (for finite $N$, the only two orthogonal atomic coherent states~(\ref{eq:atomic}) are $|0,\phi_1\rangle$ and $|\pi,\phi_2\rangle$). In a two-dimensional projection (cf.\ Fig~\ref{fig:ent}.c) such a state is a bimodal distribution (for $N\to\infty$: two delta-peaks).

To numerically identify if a given wave-function $|\psi\rangle$ is in a mesoscopic
superposition, we start by searching the atomic coherent state~$|\theta_1,\phi_1\rangle$ for
which~$|\langle\psi|\theta,\phi\rangle|^2$ reaches its maximum, $m_1$. Around
($\theta_1$,$\phi_1$) we define the set R1 by
$|\langle\theta,\phi|\theta_1,\phi_1\rangle|^2>10^{-3}$ (cf.\ Fig~\ref{fig:ent}.c). As both
parts of the mesoscopic superposition~(\ref{eq:super}) should 
hardly overlap, the second maximum $m_2=|\langle\psi|\theta_2,\phi_2\rangle|^2$ is searched
outside the set~R1. The set~R2 is
defined analogously to R1 by $|\langle\theta,\phi|\theta_2,\phi_2\rangle|^2>10^{-3}$. 
The fidelity $|\langle\psi |\psi_{\rm sp}\rangle|^2$ still is a function of $\gamma$, taking
its maximum and excluding large overlaps ($R1 \cap R2 \neq \emptyset$) yields:
\begin{equation}
\label{eq:fiddef}
p_{\rm fid} = \left\{\begin{array}{ll}
 0&:\; \rm R1\; and\; R2\; overlap \\
\frac12\left(\sqrt{m_1}+\sqrt{m_2}\right)^2&:\;\rm else
\end{array}\right.\;.
\end{equation}
Yet, this only indicates entanglement if $p_{\rm fid}>0.5$. With
\begin{equation}
\label{eq:ent} 
\sigma_{\rm ent} = \frac{m_2}{m_1}p_{\rm fid}\;,\quad \sigma_{\rm ent}\le p_{\rm fid}
\end{equation}
even values of $\sigma_{\rm ent}\lessapprox 0.5$ can identify mesoscopic superpositions (Fig.~\ref{fig:ent}.c).
In Fig.~\ref{fig:ent}.a, the maximum value of entanglement (evaluated at  $\tau=5$ and $10$) is plotted as a function of the initial condition~($z_0$, $\phi_0$):
within the chaotic regime (left), entanglement generation happens on faster time-scales than in the regular regime (right); for longer time-scales (Fig.~\ref{fig:ent}.b) the entanglement in the entire chaotic regime is more pronounced. It reaches particularly high values near initial conditions with large differences in the time-averaged population imbalances (Fig.~\ref{fig:nlssgl}.a). We obtained qualitatively similar results also for other values of driving amplitude and interaction.

To conclude, generation of mesoscopic entangled states can be a signature of quantum chaos for a BEC in a 
 periodically driven double well potential. We investigated the driving near multi-``photon'' tunneling resonances~\cite{EckardtEtAl05} which were recently observed experimentally for a BEC in an optical lattice~\cite{SiasEtAl08}. While decoherence can lead to a ``chaotic'' behavior similar to the predictions of the Gross-Pitaevskii equation, the differences between quantum dynamics and mean-field dynamics are due to the emergence of mesoscopic superpositions. If the mean-field dynamics are chaotic, the entanglement generation is accelerated and its values are enhanced.

\acknowledgments

We thank H.~P.~Breuer, Y.~Castin, A.~Eckardt and M.~Holthaus for insightful 
discussions.
Funding by the EU is gratefully 
acknowledged (CW: contract
 MEIF-CT-2006-038407; NT: contract 
MEST-CT-2005-019755).


\begin{thebibliography}{36}
\bibitem[{foo()}]{footnoteCWNT08}
\bibinfo{note}{A higher $\alpha$ leads to a larger statistical spread. $\alpha$
  includes the influence of the environment; it would, e.g., be larger for
  larger BECs.}
\end{thebibliography}


\end{document}